\documentclass[10pt,letterpaper]{article}
\usepackage[top=0.85in,left=2.75in,footskip=0.75in]{geometry}

\usepackage{amsmath,amssymb}

\usepackage{changepage}

\usepackage{textcomp,marvosym}

\usepackage{cite}

\usepackage{nameref,hyperref}

\usepackage[right]{lineno}

\usepackage[nopatch=eqnum]{microtype}
\DisableLigatures[f]{encoding = *, family = * }

\usepackage[table]{xcolor}

\usepackage{array}

\newcolumntype{+}{!{\vrule width 2pt}}

\newlength\savedwidth

\newcommand\thickhline{\noalign{\global\savedwidth\arrayrulewidth\global\arrayrulewidth 2pt}%
\hline
\noalign{\global\arrayrulewidth\savedwidth}}

\raggedright
\usepackage[aboveskip=1pt,labelfont=bf,labelsep=period,justification=raggedright,singlelinecheck=off]{caption}

\makeatletter
\renewcommand{\@biblabel}[1]{\quad#1.}
\makeatother

\usepackage{float}

\usepackage{lastpage,fancyhdr,graphicx}
\usepackage{epstopdf}
\fancyheadoffset[L]{2.25in}
\fancyfootoffset[L]{2.25in}
\begin{document}
\vspace*{0.2in}

\begin{flushleft}
{\Large
\textbf{Count data modeling and forecasting of malaria incidence using generalized time series regression} % Please use "sentence case" for title and headings (capitalize only the first word in a title (or heading), the first word in a subtitle (or subheading), and any proper nouns).
}
\newline
\\
Adithya B. Somaraj\textsuperscript{1},
Praveen D. Chougale\textsuperscript{1, 2},
Usha Ananthakumar\textsuperscript{1, 2, 3},
Karthika M. Satyanarayanan\textsuperscript{1}
\\
\bigskip
\textbf{1} National Disease Modeling Consortium, Indian Institute of Technology Bombay, Mumbai, Maharashtra, India
\\
\textbf{2} Koita Centre for Digital Health, Indian Institute of Technology Bombay, Mumbai, Maharashtra, India
\\
\textbf{3} Shailesh J. Mehta School of Management,  Indian Institute of Technology Bombay, Mumbai, Maharashtra, India
\\
\bigskip

% Insert additional author notes using the symbols described below. Insert symbol callouts after author names as necessary.
% 
% Remove or comment out the author notes below if they aren't used.
%
% Primary Equal Contribution Note
%\Yinyang These authors contributed equally to this work.

% Additional Equal Contribution Note
% Also use this double-dagger symbol for special authorship notes, such as senior authorship.
%\ddag These authors also contributed equally to this work.

%\textcurrency Current Address: Dept/Program/Center, Institution Name, City, State, Country % change symbol to "\textcurrency a" if more than one current address note
% \textcurrency b Insert second current address 
% \textcurrency c Insert third current address


% Group/Consortium Author Note
%\textpilcrow Membership list can be found in the Acknowledgments section.

% Use the asterisk to denote corresponding authorship and provide email address in note below.
* usha@iitb.ac.in

\end{flushleft}

% For PLOS Medicine research article authors, please structure your abstract
% with "Background", "Method and Findings" and "Conclusion" sections per
% journal requirements.

% For PLOS Neglected Tropical Diseases research article authors, please
% structure your abstract with "Background", "Methodology", "Findings", and
% "Conclusion" sections per journal requirements.
%
\section*{Abstract}

Malaria remains a major public health concern in many urban regions of India, where timely prediction of malaria incidence is essential for effective surveillance and resource allocation. This study examines count data approaches for understanding and predicting malaria incidence in the Mumbai region. The analysis used monthly \textit{Plasmodium vivax} surveillance data from the Health Management Information System (HMIS) collected between 2012 and 2019, together with meteorological variables. Initial Poisson regression models suggested strong associations between malaria incidence and environmental factors; however, diagnostic assessment revealed substantial overdispersion, indicating that the Poisson model did not adequately capture the data's variability. Negative binomial regression provided a better representation of the data and indicated that seasonal effects were more strongly associated with malaria incidence than individual climatic covariates. Residual analyses further identified significant serial dependence not captured by baseline regression models. To address this limitation, a Generalized Linear Autoregressive Moving Average (GLARMA) framework was implemented to model temporal correlation explicitly. Forecasts were generated using simulation-based methods and evaluated through rolling time series cross-validation. The GLARMA Negative binomial model consistently demonstrated superior predictive performance and greater predictive stability than competing regression and time series approaches. These findings highlight the importance of jointly accounting for overdispersion and serial dependence in malaria surveillance data and demonstrate the value of count time series models for supporting early warning systems in urban settings.

% Please keep the Author Summary between 150 and 200 words. Use first person.
% PLOS ONE, PLOS Biology, PLOS Global Public Health, PLOS Mental Health, and PLOS Water authors please skip this step. Author Summary is not valid for submissions to these journals.

% For PLOS Medicine authors, please structure your author summary with answers to the following questions:
% Why was this study done?
% What did the researchers do and find?
% What do these findings mean?
%
% \section*{Author summary}
% Lorem ipsum dolor sit amet, consectetur adipiscing elit. Curabitur eget porta erat. Morbi consectetur est vel gravida pretium. Suspendisse ut dui eu ante cursus gravida non sed sem. Nullam sapien tellus, commodo id velit id, eleifend volutpat quam. Phasellus mauris velit, dapibus finibus elementum vel, pulvinar non tellus. Nunc pellentesque pretium diam, quis maximus dolor faucibus id. Nunc convallis sodales ante, ut ullamcorper est egestas vitae. Nam sit amet enim ultrices, ultrices elit pulvinar, volutpat risus.

\nolinenumbers

% Use "Eq" instead of "Equation" for equation citations.

\section*{Introduction}

Malaria remains an important public health challenge in many tropical and subtropical regions despite substantial progress in disease control over recent decades. In India, routine surveillance systems generate monthly malaria incidence data that guide healthcare planning, vector control activities, and resource allocation. These surveillance records provide an opportunity to develop forecasting models capable of anticipating disease outbreaks and informing public health interventions~\cite{thomson2006,who2024}. However, modeling malaria incidence presents several statistical challenges because surveillance records consist of discrete count data that often exhibit overdispersion and temporal dependence. In rapidly urbanizing settings such as Mumbai, transmission dynamics are further influenced by population density, environmental heterogeneity, monsoon rainfall, and seasonal fluctuations, resulting in complex temporal patterns~\cite{zeileis2008,battle2019,white2011,sharma2009}.

Generalized Linear Models (GLMs) are widely used to analyze count data in epidemiological research. Poisson regression is commonly adopted as a baseline approach, while negative binomial regression extends the framework by accommodating overdispersion, a phenomenon in which the variance exceeds the mean~\cite{Nelder1972,hilbe2011,zeileis2008}. In Mumbai malaria surveillance, Poisson and negative binomial GLMs have been applied to monthly \textit{Plasmodium vivax} counts to compare interpretable regression-based models with machine learning alternatives~\cite{chougale2025}. Although these models capture important distributional properties of count data, they generally assume conditional independence across observations. For infectious diseases such as malaria, this assumption is often unrealistic because disease incidence is influenced by previous transmission activity, climatic persistence, vector ecology, and relapse mechanisms associated with \textit{Plasmodium vivax} infections~\cite{battle2019,white2011,briet2013}. Ignoring such temporal dependence can result in biased inference, underestimated uncertainty, and reduced forecasting accuracy~\cite{kedem2002,briet2013}.

Observation-driven approaches such as the Generalized Linear Autoregressive Moving Average (GLARMA) model extend conventional GLMs by incorporating autoregressive dependence directly into the modeling framework~\cite{Li1994,davis2003,dunsmuir2015}. GLARMA retains the interpretability of regression-based methods while accounting for serial correlation in non-Gaussian count data, and has been applied to respiratory and environmental count time series~\cite{Camara2025Applied,camara2025}. GLARMA models have previously been applied to malaria counts from a major tertiary hospital in Mumbai, where explicitly modeling temporal dependence improved forecasting relative to other regression models~\cite{mukhopadhyay2019}. However, these studies do not address the risk of error accumulation when forecasting with GLARMA, nor do they evaluate the stability of forecasts across and within multiple horizons.

To address these gaps, we model monthly malaria incidence in the Mumbai region (both Mumbai Suburban and 228
Mumbai City districts) from 2012 to 2019 with rigorous in-sample diagnostics, starting with GLMs to identify the appropriate distribution and extending it to the GLARMA framework to correct for the underlying temporal autocorrelation. Furthermore, we use the median of conditional means of the simulated forecasts to mitigate the risk of inflation and systematically compare forecasts across four horizons using time series cross-validation against Gaussian ARIMA benchmarks to ensure the robustness of GLARMA forecasts.

\section*{Methods}
\subsection*{Generalized linear models}

The Generalized Linear Models (GLMs) are a class of models generalizing the ordinary linear regression for data where the dependent variable follows non-normal distributions like Poisson, negative binomial, gamma, etc., and are fitted based on likelihood \cite{Nelder1972}. These distributions have a probability density function (PDF) or a probability mass function (PMF) of the form:
\begin{align}
\label{eq:expdispfam}
    f(y_i;\theta_i,\phi) = \textrm{exp}\left\{\frac{y_i\theta_i-b(\theta_i)}{a(\phi)} +  c(y_i, \phi)\right\}
\end{align}
for natural parameter $\theta_i$, dispersion parameter $\phi$ and arbitary functions $a(.)$, $b(.)$ and $c(.)$. By Eq (\ref{eq:expdispfam}), it also follows that the mean and variance of this family of distributions are
\begin{align}
\label{eq:mean} 
    \mu_i = E(y_i) &= b'(\theta_i) \phantom{\mu_i = E(y_i)} \\
\label{eq:var} 
    \phantom{\mu_i = E(y_i)} V(y_i) &= b''(\theta_i)a(\phi) \phantom{\mu_i = E(y_i)}
\end{align}
where $b'(\theta_i)$ and $b''(\theta_i)$ represent the first and second derivatives of the cumulant function $b(\theta_i)$ with respect to the natural parameter, which mathematically govern the mean and the specific mean-variance relationship characteristic of the chosen distribution.

GLMs connect the expected value of the dependent variable to a linear combination of the independent variables and coefficients using a link function $g(.)$ as
\begin{align}
\label{eq:glm}
    g(E(\mathbf{Y})) = \boldsymbol{\eta} &= \mathbf{X}\boldsymbol{\beta}
\end{align}
where $E(\mathbf{Y})$ is the expected value of the dependent variable, $\mathbf{X}$ is the matrix containing all the independent variables and $\boldsymbol{\beta}$ is the coeffcient vector. We use some special cases of GLMs, Poisson and negative binomial, to model data in the form of counts, i.e., discrete variables \cite{BelCountData, zeileis2008}.

\subsubsection*{Poisson regression}
The Poisson distribution is the simplest distribution used to model non-binary discrete variables. It has the probability mass function
\begin{align}
\label{eq:poiss}
    f(y_i;\mu_i) = \frac{e^{-\mu_i}\mu_i^{y_i}}{y_i!} &= \textrm{exp}\left\{y_i\log\mu_i - \mu_i - \log(y_i!)\right\}
\end{align}

Since we can rearrange Eq (\ref{eq:poiss}) such that the natural parameter $\theta_i = \log\mu_i$, $b(\theta_i) = e^{\theta_i}$, $a(\phi)=1$ and $c(y_i, \phi) = 1$, it is of the family of Eq (\ref{eq:expdispfam}) and a special case of the GLM framework. From Eq (\ref{eq:mean}) and Eq (\ref{eq:var}),
\begin{align}
\label{eq:poissmean} 
    E(y_i) &= \mu_i \\
\label{eq:poissvar} 
   V(y_i) &= \mu_i
\end{align}

The canonical link is $g(\mu_i) = \log\mu_i$, resulting in a log-linear relationship between mean and linear predictor, and the dispersion is fixed at $c(\phi) = 1$. The Poisson model successfully describes the mean $\mu_i$ but often underestimates the variance of the data, rendering all the model-based statistical tests invalid.

\subsubsection*{Negative binomial regression}
Poisson distribution assumes that the mean and the variance are the same, but there are many cases where the variance exceeds the mean (overdispersion), such as the malaria data we have used \cite{Ghosh2025}. One way to model count data with overdispersion is to use the negative binomial distribution. The probability mass function of $y_i$ following a negative binomial distribution is
\begin{align}
\label{eq:negbinpmf}
    f(y_i;\mu_i,k) = \frac{\Gamma(y_i+k)}{\Gamma(k)\Gamma(y_i+1)}\left(\frac{\mu_i}{\mu_i+k}\right)^{y_i}\left(\frac{k}{\mu_i+k}\right)^{k}
\end{align}
where $\Gamma(.)$ is the gamma function and $k$ is the shape parameter. For every fixed $k$, this is of type Eq(\ref{eq:expdispfam}) and hence, from Eq (\ref{eq:mean}) and Eq (\ref{eq:var}),
\begin{align}
\label{eq:negbinmean} 
    E(y_i) &= \mu_i \\
\label{eq:negbinvar} 
   V(y_i) &= \mu_i + \frac{\mu_i}{k^2}
\end{align}

\subsection*{The GLARMA model}
Generalized linear models (GLMs) for time series are widely used in various applications to model the relationship between a non-Gaussian dependent variable (such as count or binary data) and one or more independent variables over time. One common issue encountered in such models is autocorrelation in the data, which violates the assumption of independence of observations in standard GLM estimation. Ignoring this autocorrelation can lead to inefficient estimates and misleading inferences \cite{Box2015, Tsay2010, Ghosh2025}. 

The Generalized Linear Autoregressive Moving-Average (GLARMA) model is a subclass of observation-driven generalized state space models for non-Gaussian time series \cite{Li1994, davis2003, dunsmuir2015}. The state process is related linearly to the regressor variables and non-linearly to the past values of the observed process. 

The response series is denoted by $\{Y_t\}, t \in \mathbb{Z}$ which is associated with $k$ independent variables denoted by $\{\mathbf{X}_{i,t}\},\ 1 \leq i \leq k$. We let $\mathcal{F}_{t-1}=\sigma \{Y_s,s \leq t-1;\mathbf{X}_{i,s},1 \leq i \leq k,s \leq t \}$ be the past responses and the past and present of regressors relative to time $t$. Conditional on $\mathcal{F}_{t-1}$, the distribution of $Y_t$ is independent and assumed to belong to the exponential family defined previously in Eq (\ref{eq:expdispfam}). In this dynamic context, the canonical parameter is denoted by $W_t$, which summarizes the information in $\mathcal{F}_{t-1}$. Following Eq (\ref{eq:mean}) and Eq (\ref{eq:var}), the conditional mean and variance are given by $\mu_t = E(Y_t|\mathcal{F}_{t-1}) = b'(W_t)$ and $\sigma^2_t = V(Y_t|\mathcal{F}_{t-1})=b''(W_t)a(\phi)$, respectively. The canonical link function $g(\cdot)$ is used to relate $\mu_t$ to $W_t$ such that $W_t = g(\mu_t)$. 

Expanding upon the standard GLM linear predictor specified in Eq (\ref{eq:glm}), the general form of $W_t$ in the GLARMA framework integrates an autoregressive state process:
\begin{align}
\label{eq:glarma_wt}
    W_t = \mathbf{X}^T_t\boldsymbol{\beta} + Z_t, 
\end{align}
where $\mathbf{X}_t = (1,\ X_{1,t},\ X_{2,t},...,X_{k,t})^T$ are the independent variables of $k+1$ dimension and $\boldsymbol{\beta} = (\beta_0,\ \beta_1,...,\beta_k)^T$ are the regression coefficients. $\{Z_t\}$ is the noise process introducing a serial dependence on the observations, without which Eq (\ref{eq:glarma_wt}) would reduce to the standard independent GLM. It is given by:
\begin{align}
\label{eq:glarma_zt}
     Z_t = \sum_{i=1}^{\infty}\ \gamma_i \ e_{t-i},
\end{align}
where the infinite moving average weights $\gamma_i$ can take the form of an autoregressive moving-average (ARMA) filter:
\begin{align}
\label{eq:glarma_arma}
     \sum_{i=1}^{\infty}\ \gamma_i \ z^i = \frac{\theta (z)}{\phi(z)} - 1, \qquad |z| \leq 1.
\end{align}

The autoregressive and moving average components $\phi(z) = (1-\phi_1z-...-\phi_pz^p)$ and $\theta(z) = (1+\theta_1z +...+\theta_qz^q)$ are polynomials with roots outside the unit circle and have no common zeroes. The model given by Eq (\ref{eq:glarma_wt}), Eq (\ref{eq:glarma_zt}), and Eq (\ref{eq:glarma_arma}) is the GLARMA$(p,q)$ model. The process $Z_t$ can be computed recursively as:
\begin{align}
\label{eq:glarma_recursive}
     Z_t = \phi_1(Z_{t-1} + e_{t-1}) + ... + \phi_p(Z_{t-p} + e_{t-p})+\theta_1e_{t-1}+...+\theta_qe_{t-q}.
\end{align}
The standard predictive residuals $\{e_t\}$ driving the state process in Eq (\ref{eq:glarma_zt}) and Eq (\ref{eq:glarma_recursive}) are defined as:
\begin{align}
\label{eq:glarma_resid}
     e_t = \frac{Y_t -\mu_t}{\sigma_t}.
\end{align}

\subsection*{Time series decomposition}

Time series count data modeled via GLMs can be conceptualized by decomposing the systematic component of the linear predictor. Recall from Eq (\ref{eq:glarma_wt}) that the canonical parameter is $W_t = \mathbf{X}^T_t\boldsymbol{\beta} + Z_t$. To understand the underlying patterns within the non-Gaussian data, the design matrix $\mathbf{X}_t$ and the coefficient vector $\boldsymbol{\beta}$ can be partitioned into primary structural elements: trend ($\mathbf{X}_{tr,t}^T\boldsymbol{\beta}_{tr}$), seasonality ($\mathbf{X}_{seas,t}^T\boldsymbol{\beta}_{seas}$), and external covariates ($\mathbf{X}_{co,t}^T\boldsymbol{\beta}_{co}$). $\mathbf{X}^T_t\boldsymbol{\beta}$ is therefore the sum of these sub-components:
\begin{align}
\label{eq:decomposition}
    \mathbf{X}^T_t\boldsymbol{\beta} = \beta_0 + \mathbf{X}_{tr,t}^T\boldsymbol{\beta}_{tr} + \mathbf{X}_{seas,t}^T\boldsymbol{\beta}_{seas} + \mathbf{X}_{co,t}^T\boldsymbol{\beta}_{co}
\end{align}
where $\beta_0$ is the global baseline intercept.

\subsubsection*{Trend}
The trend component encapsulates the long-term movement in the data on the link-function scale, often driven by underlying factors such as population dynamics or gradual environmental changes. It can be modeled as a linear predictor:
\begin{align*}
    \mathbf{X}_{tr,t}^T\boldsymbol{\beta}_{tr} = \beta_{tr,1} t
\end{align*}
where $t$ is the continuous time index.

\subsubsection*{Seasonality} 
Seasonality refers to the periodicity within a time series, typically driven by weather cycles in epidemiological data \cite{Ghosh2025}. Since seasonality in monthly data is generally characterized by a 12-month frequency, it is incorporated using dummy variables:
\begin{align*}
    \mathbf{X}_{seas,t}^T\boldsymbol{\beta}_{seas} = \beta_{seas,1} D_{1,t} + \beta_{seas,2} D_{2,t} + \dots + \beta_{seas,11} D_{11,t}
\end{align*}
where $D_{i,t}$ represents the months of the year, with one month acting as the reference intercept to avoid perfect collinearity.

\subsubsection*{Covariates}
In epidemiological applications, the dependent variable is heavily influenced by external environmental factors that change over time. These are included in the model as a subset of continuous covariates, $\mathbf{X}_{co,t}$, with $\boldsymbol{\beta}_{co}$ representing the vector of coefficients that estimate the independent effect of each specific predictor.

\subsubsection*{The latent autoregressive process}
In standard GLMs, the strict assumption is that observations are independent given the structural predictors defined above. However, in infectious disease time series, unexplained transmission momentum often remains, resulting in autocorrelated predictive residuals. 

In the GLARMA framework, rather than adding an error term directly to the discrete observation, we explicitly model the latent autocorrelated process $Z_t$. To ensure computational stability and avoid overparameterization within the non-Gaussian framework, we restrict our latent process to an Autoregressive AR($p$) structure. By setting the moving-average parameters ($q$) to zero in the general GLARMA recursion from Eq(\ref{eq:glarma_recursive}), the observation-driven latent state reduces to:
\begin{align}
\label{eq:ar_latent}
    Z_t = \sum_{l=1}^p \phi_l (Z_{t-l} + e_{t-l})
\end{align}
where $\phi_l$ are the autoregressive parameters and $e_{t-l}$ are the past standard predictive residuals (as defined in Eq (\ref{eq:glarma_resid})) acting as the driving white noise. By appending this explicitly modeled AR process to the decomposed systematic matrix $\mathbf{X}^T_t\boldsymbol{\beta}$, the final GLARMA model accounts for both structural drivers and temporal dependence.

\subsection*{Model evaluation and forecasting}

\subsubsection*{Residual diagnostics}

Residuals and plots of residuals are crucial in the evaluation of statistical models. In normal linear regression, the residuals are assumed to be normally distributed with constant variance (homoscedasticity). In non-normal regression scenarios, such as when the dependent variable is discrete, the standard residuals do not satisfy these assumptions. They align in almost parallel curves, rendering visual inspection and standard diagnostic tests largely uninformative. 

We used randomized quantile residuals to assess the statistical validity of the count regression models \cite{dunn1996, SadeghpourRQR}. This method leverages the exact distribution of the fitted model via the Probability Integral Transform. For discrete data, it calculates the mathematical cumulative probability boundaries of the observed data and applies a single randomization step to produce continuous, standard normal residuals $\mathcal{N}(0,1)$ \cite{dunn1996}. These residuals are then mathematically projected onto a standard uniform distribution $\mathcal{U}(0,1)$ using the standard normal cumulative distribution function \cite{Diebold1998}, so that visual interpretation and outlier detection are made straightforward \cite{hartig2022}. Residual diagnostics were performed by analyzing two primary visualizations \cite{hartig2022}: 
\paragraph{Quantile-quantile (Q-Q) plot}
This plot compares the empirical distribution of the uniform-projected residuals to the expected theoretical uniform distribution. Structural deviation from the expected distribution is formally assessed using the Kolmogorov-Smirnov (KS) test, while overdispersion is independently tested using the $\chi^2$ test on the variance of the underlying normal residuals.
\paragraph{Residuals vs. predicted plot}
This plot visualizes the uniform residuals against the rank-transformed predicted values. For a correctly specified model, the residuals should be randomly scattered across the vertical axis. A LOESS smoothing curve was overlaid to detect systematic nonlinearities in quantiles across the predictive range, and extreme outliers (residuals mathematically forced to $0$ or $1$) were explicitly highlighted.

\subsubsection*{Forecasting via Monte Carlo simulation}

Due to the non-linear link function $g(\cdot)$ (since $E[g^{-1}(\eta)] \neq g^{-1}(E[\eta])$) and the reliance of future latent states on unobserved future responses in GLARMA, $h$-step-ahead out-of-sample forecasts cannot rely on simple formulas. Therefore, predictive distributions are generated using Monte Carlo simulation \cite{dunsmuir2015}. For $M$ simulated future paths and forecast horizon $h$, at each future time step $t \in \{n+1, \dots, n+h\}$ and for each iteration $m \in \{1, \dots, M\}$, the process operates sequentially:
\begin{align}
\eta_{t}^{(m)} &= \mathbf{X}_{t}^T\boldsymbol{\beta} + Z_{t}^{(m)} \\
\mu_{t}^{(m)} &= g^{-1}\left( \eta_{t}^{(m)} \right) \\
y_{t}^{(m)} &\sim f\left(y_t; \mu_{t}^{(m)}, k\right) \\
Z_{t+1}^{(m)} &= \sum_{l=1}^p \phi_l \left( Z_{t+1-l}^{(m)} + e_{t+1-l}^{(m)} \right)
\end{align}
Here, $f(\cdot)$ is the assumed PMF, and $e_{t+1-l}^{(m)}$ represents the simulated predictive residuals driving the state process. 

This multi-step simulation process comes with the risk of   ``explosive paths," where drawing a randomly large outlier when simulating future responses $y_{t}^{(m)}$ can cause the subsequent predictive residual and autoregressive state $Z_{t+1}^{(m)}$ to be inflated, creating a compounding upward spiral for that specific simulated path. This is most pronounced in heavy-tailed distributions like negative binomial, where the variance grows quadratically with the mean as shown in Eq(\ref{eq:negbinvar}), but is also a structural risk even in Poisson. 

Instead of taking the arithmetic mean of the raw simulated responses, which would be skewed upwards by the outliers, we extract the underlying conditional state $\mu_{t}^{(m)}$ and compute the final point forecast using the median for all $M$ simulations. 
This provides a mathematically safer estimate shielded from extreme outlier paths, ensuring stable multi-step forecasts.
\begin{align}
\hat{\mu}_{t} &= \text{Median} \left( \{ \mu_{t}^{(1)}, \mu_{t}^{(2)}, \dots, \mu_{t}^{(M)} \} \right)
\end{align}

\subsubsection*{Forecast accuracy}

Testing a time series model on unseen data is crucial for evaluating its predictive performance \cite{Ghosh2025}. This ensures the model's generalizability to data not used during model fitting. This evaluation is typically conducted using a variety of accuracy metrics, with the Root Mean Squared Error (RMSE) and the Mean Absolute Percentage Error (MAPE) being the most common. 

Suppose we use the historical dataset $\mathcal{D}_n = \{y_1, y_2, \dots , y_n\}$ to estimate the parameters of the model, and let the test values be denoted by $\mathcal{T}_h = \{y_{n+1}, y_{n+2}, \dots , y_{n+h}\}$. If $\hat{\mu}_{n+t}$ denotes the point forecast for time $n+t$, then RMSE and MAPE are given by 
\begin{align}
\text{RMSE} = \sqrt{\frac{1}{h} \sum_{t=1}^{h} \left( y_{n+t} - \hat{\mu}_{n+t} \right)^2}
\end{align}

\begin{align}
\text{MAPE} = \frac{100}{h} \sum_{t=1}^{h} \left| \frac{y_{n+t} - \hat{\mu}_{n+t}}{y_{n+t}} \right|
\end{align}
A combination of residual analysis and out-of-sample prediction provides a comprehensive assessment of model performance and is essential for validating any time series model.

\subsubsection*{Time series cross-validation}

To dynamically evaluate the stability of the models' forecasts, we used a time series cross-validation approach often called ``evaluation on a rolling forecasting origin" \cite{Hyndman2021}. To this end, the data is partitioned into training and testing sets in a rolling manner rather than relying on a single train-test split. The model is re-estimated, and out-of-sample forecasts are generated for a specified number of steps ahead (e.g., $h = 3$ steps) for each rolling window. The error metrics from these rolling forecasts can then be analyzed to see how the model's predictive accuracy holds up over different training windows.

\section*{Methodology}
\subsection*{Data and preprocessing}
For this study, we used data from the Mumbai region for the period January, 2012 to December, 2019. Monthly records of total malaria case counts due to the $Plasmodium Vivax$ parasite, monthly meteorological data, and demographic data were compiled from various sources. The malaria case counts were obtained from the Health Management Information System (HMIS), a web-based data portal established by the Ministry of Health \& Family Welfare (MoHFW), Government of India, to monitor health programs \cite{MoHFW_HMIS}. The three missing values in this malaria data were imputed using linear interpolation and rounded to the nearest integer. The period for this study was chosen based on the availability of data on the HMIS portal. The meteorological data obtained from the Climate Data Service Portal, an integrated platform that provides weather and climate services from the Indian Meteorological Department, Government of India, were used as covariates. We selected monthly rainfall (in cm), mean maximum temperature (in Celcius), and mean duration of sunshine (in hours) as the covariates after performing VIF-based elimination of variables to address multicollinearity. Estimated population projections by the International Institute for Population Sciences \cite{Dhar2022} were used as the offset vector for GLARMA.

\subsubsection*{In-sample modeling}

We started by estimating baseline Poisson and negative binomial GLMs as defined in Eq (\ref{eq:decomposition}), operating under the assumption that the observed case counts $y_t$ are mutually independent, conditional on the structural predictors. To validate this assumption and assess overall goodness-of-fit, we visually vetted their randomized quantile residuals using the Q-Q and residual vs. predicted plots, and statistically using the Kolmogorov-Smirnov (KS) test and the $\chi^2$ dispersion test. After using the Durbin-Watson test and Autocorrelation Function (ACF) plots to test for temporal autocorrelation, we extended our modeling to the GLARMA framework using Eq. (\ref{eq:ar_latent}) and verified its adequacy using the same autocorrelation tests.

\subsubsection*{Benchmarking out-of-sample forecasts}

We compared ex-post forecasts (forecasts where predictor information is available for the forecast period) \cite{Hyndman2021} from four different regression models to establish the best predictive framework. All candidate models share the same $\mathbf{X}_t^T \boldsymbol{\beta}$ (defined previously in Eq (\ref{eq:decomposition})), incorporating trend, seasonality and covariates. 

The first two models are standard Gaussian frameworks with autoregressive (AR) error corrections, and the latter two use GLARMA. We used an AR order ($p$) of 2  for all models to ensure consistency.
\paragraph{LM + ARIMA}
This model fits the case counts using standard ordinary least squares (OLS) assumptions, with an autoregressive error term to handle temporal dependencies, and acts as a baseline.
\begin{align}
y_t &= \mathbf{X}_t^T \boldsymbol{\beta} + \epsilon_t \\
\epsilon_t &= \sum_{l=1}^p \phi_l \epsilon_{t-l} + u_t
\end{align}
where $u_t$ is Gaussian white noise. The forecast is generated directly as $\hat{y}_t = \mathbf{X}_t^T \hat{\boldsymbol{\beta}} + \hat{\epsilon}_t$.
\paragraph{Log-LM + ARIMA}
This model applies a logarithmic transformation to the case counts to address potential heteroscedasticity and exponential growth.
\begin{align}
\log(y_t) &= \mathbf{X}_t^T \boldsymbol{\beta} + \epsilon_t \\
\epsilon_t &= \sum_{l=1}^p \phi_l \epsilon_{t-l} + u_t
\end{align}
Because forecasting requires transforming predictions back to the original count scale, we must account for the residuals' variance to avoid underestimation. Assuming $\log(y_t)$ follows a normal distribution $\mathcal{N}(\mu_t, \sigma^2)$, the expected value on the original scale is corrected using the properties of the log-normal distribution [26]:
\begin{align}
\hat{y}_t = \exp\left( \mathbf{X}_t^T \hat{\boldsymbol{\beta}} + \hat{\epsilon}_t + 0.5\hat{\sigma}^2 \right)
\end{align}
\paragraph{GLARMA Poisson}
Transitioning to GLARMA, this model treats the response variable as a Poisson distribution, driven by an autoregressive latent state. The variance is assumed to be strictly equal to the mean.
\begin{align}
y_t | \mathcal{F}_{t-1} &\sim \text{Poisson}(\mu_t) \\
\eta_t &= \mathbf{X}_t^T \boldsymbol{\beta} + Z_t \\
\log(\mu_t) &= \eta_t \\
Z_t &= \sum_{l=1}^p \phi_l (Z_{t-l} + e_{t-l})
\end{align}
where $\eta_t$ is the full linear predictor, $Z_t$ is the latent AR state, and $e_{t-l}$ are the past predictive residuals.
\paragraph{GLARMA Negative binomial}
To account for the overdispersion often present in epidemiological data, we use GLARMA with negative binomial assumption. 
\begin{align}
y_t | \mathcal{F}_{t-1} &\sim \text{NB}(\mu_t, k) \\
\eta_t &= \mathbf{X}_t^T \boldsymbol{\beta} + Z_t \\
\log(\mu_t) &= \eta_t \\
Z_t &= \sum_{l=1}^p \phi_l (Z_{t-l} + e_{t-l})
\end{align}
where $\alpha$ is the estimated dispersion parameter.

 We evaluated forecast accuracy using time series cross-validation across 4 prediction windows ($h = \{3, 6, 9, 12\} $) by calculating RMSE and MAPE, and checked whether the MAPE values were stable in each prediction window by visually inspecting their boxplots.

% Results and Discussion can be combined.
\section*{Results and discussion}

\begin{figure}[htbp]
    \centering
    \includegraphics[width=0.98\textwidth]{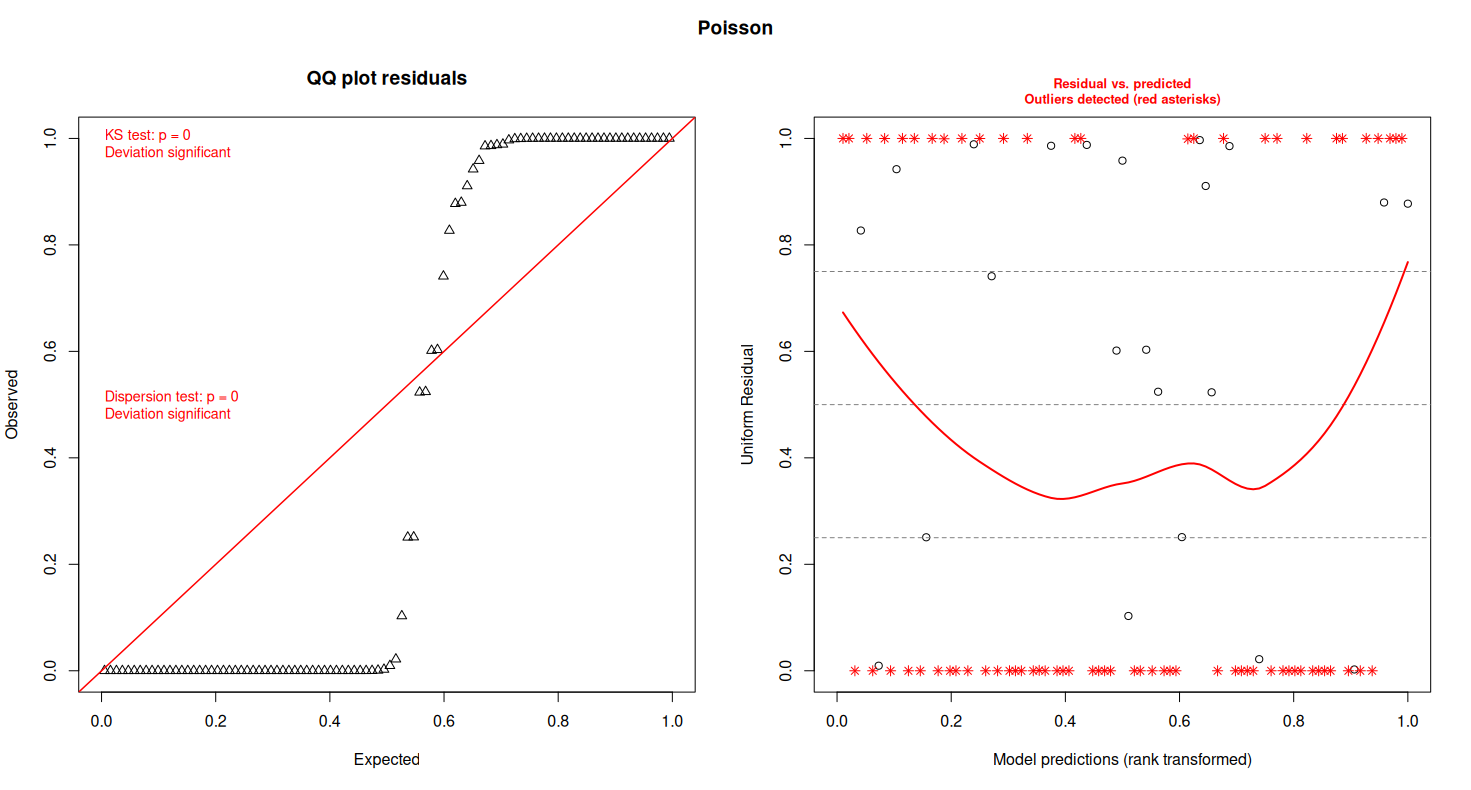}
\caption{{\bf Baseline Poisson GLM residual diagnostics.} Randomized quantile residuals showing significant deviations from the expected uniform distribution.}
    \label{fig:poiss_diagnostics}
\end{figure}

The standard Poisson regression model yielded seemingly highly significant parameter estimates across almost all predictors ($p < 0.001$), but an assessment of the model's goodness-of-fit revealed severe structural violations. The residual deviance ($12,401$) vastly exceeded the residual degrees of freedom ($80$), indicating extreme overdispersion and poor fit, which was corroborated by the Chi-square goodness-of-fit test ($p \approx 0$). The randomized quantile residuals for the Poisson model in Fig \ref{fig:poiss_diagnostics} showed large deviations from the expected uniform distribution, with the Kolmogorov-Smirnov (KS) test and the dispersion test both yielding $p$-values of approximately 0. The residuals versus predicted plot also showed many outliers, which is a classic indicator of unaddressed overdispersion. As a result, the standard errors in the Poisson model are also artificially deflated, leading to a high rate of Type I errors (false positives) in assessing covariate significance.

\begin{figure}[htbp]
    \centering
    \includegraphics[width=0.98\textwidth]{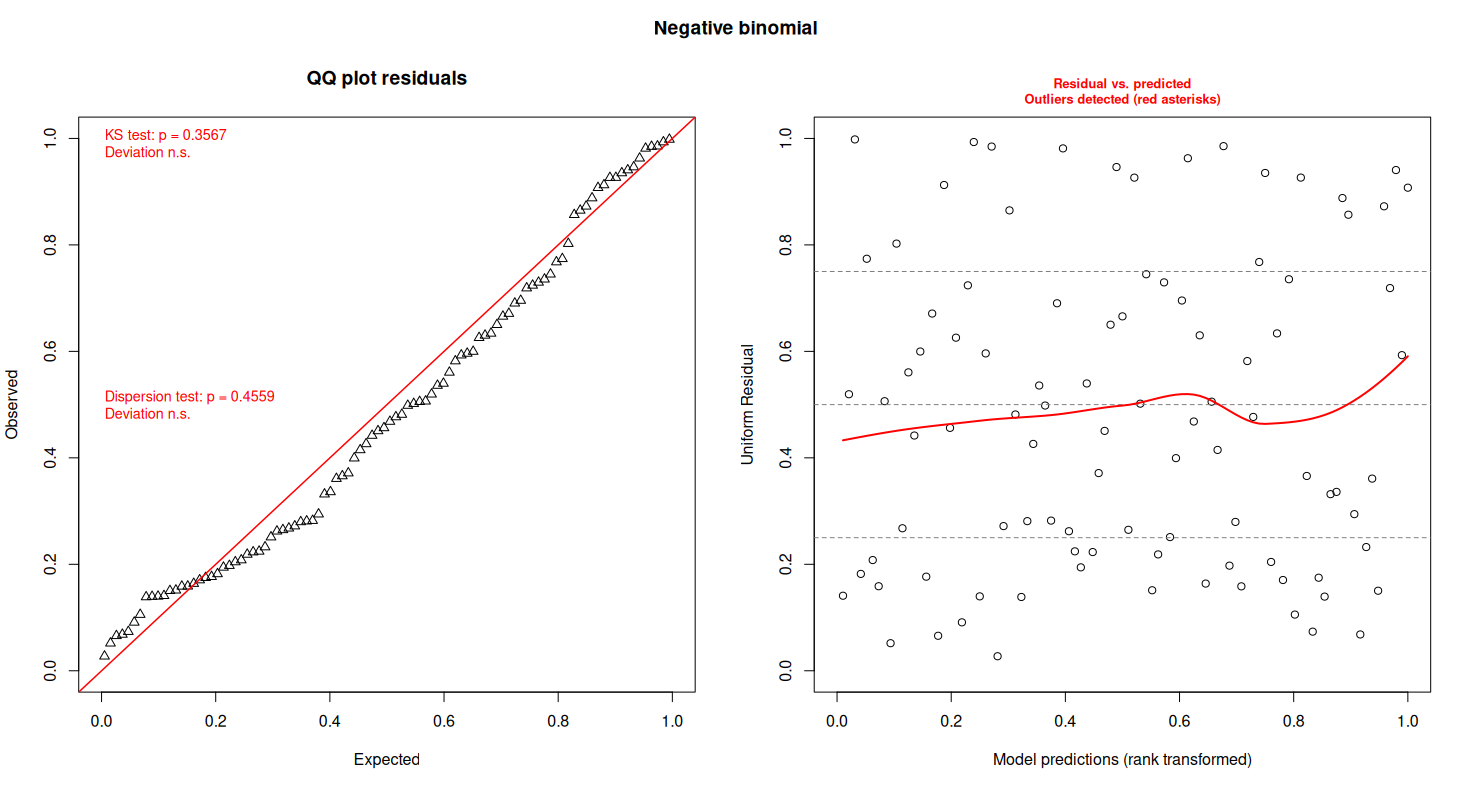}
    
    \caption{{\bf Baseline negative binomial GLM residual diagnostics.} Randomized quantile residuals perfectly aligning with the uniform distribution.}
    \label{fig:nb_diagnostics}
\end{figure}

On fitting the negative binomial (NB) GLM to correct for this overdispersion, the estimated dispersion parameter ($\alpha = 6.80$, standard error $= 0.97$) confirmed that the variance significantly exceeded the mean. The performance and diagnostic metrics for both baseline models are shown in Table \ref{table:glm_comparison}. We can see that the NB model provides a better overall model fit, reducing the AIC from $13,253$ in the Poisson model to $1,412.2$. The residual deviance ($98.30$ on $80$ degrees of freedom) also aligned closely with the expected degrees of freedom. We can see in the diagnostic plots for the NB model (Fig \ref{fig:nb_diagnostics}) that there is a near-perfect alignment along the Q-Q plot (KS test $p = 0.647$) and the randomized quantile residuals versus fitted values show a stable, homoscedastic scatter. The  dispersion test failed to reject the null hypothesis ($p = 0.456$)

\begin{table}[!ht]
\begin{adjustwidth}{-2.25in}{0in}
\centering
\caption{\textbf{In-sample performance and diagnostic metrics for baseline GLMs.} Comparison of predictive accuracy and structural goodness-of-fit between the Poisson and negative binomial models.}
\begin{tabular}{|l+c|c|}
\hline
\multicolumn{1}{|l|}{\textbf{Metric}} & \multicolumn{1}{|c|}{\textbf{Poisson GLM}} & \multicolumn{1}{|c|}{\textbf{Negative binomial GLM}} \\ \thickhline
Root Mean Squared Error (RMSE) & 368.24 & 398.94 \\ \hline
Mean Absolute Percentage Error (MAPE) & 33.35\% & 34.45\% \\ \hline
Akaike Information Criterion (AIC) & 13,253 & 1,412.2 \\ \hline
Residual Deviance & 12,401 & 98.30 \\ \hline
Residual Degrees of Freedom & 80 & 80 \\ \hline
Dispersion test ($\chi^2$) $p$-value & $<0.001$ & 0.456 \\ \hline
KS Test $p$-value & $<0.001$ & 0.647 \\ \hline
Dispersion Parameter ($\alpha$) & 1.00 (Fixed) & 6.80 (SE = 0.97) \\ \hline
\end{tabular}
\label{table:glm_comparison}
\end{adjustwidth}
\end{table}

\begin{figure}[htbp]
    \centering
    \includegraphics[width=0.51\textwidth]{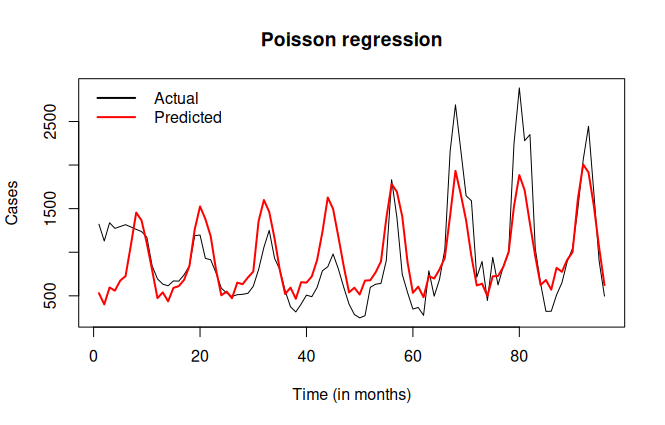}
    \hfill
    \includegraphics[width=0.473\textwidth]{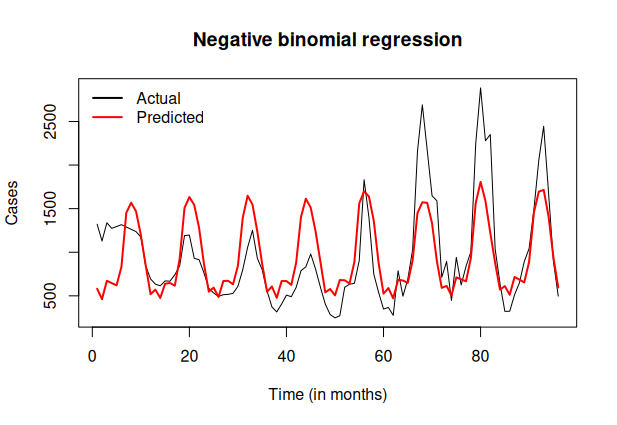}
    \caption{{\bf In-sample actual versus predicted \textit{Plasmodium vivax} cases.} (Left) Predictions from the baseline Poisson GLM. (Right) Predictions from the baseline negative binomial GLM.}
    \label{fig:ts_comparison}
\end{figure}

As expected in data with overdispersion, transitioning from the Poisson to the negative binomial distribution shows a pronounced shift in statistical inference. In the strictly constrained Poisson model, nearly all predictors, including environmental factors (maximum temperature, rainfall, and sunshine), were deemed highly statistically significant. However, once the NB model properly accounted for this, these environmental covariates lost their statistical significance ($p > 0.6$), illustrating how unaddressed overdispersion can inflate the apparent statistical significance of independent variables. Instead, the explanatory power consolidated around the late-summer and autumn seasonal dummy variables (August, September, and October; $p < 0.05$). This could be because seasonal variables capture the variance and temporal clustering more effectively.
The negative binomial model yielded an RMSE of $398.94$ and a MAPE of $34.45\%$ in-sample. This is a slight degradation compared to the Poisson model's RMSE of $368.24$ and MAPE of $33.35\%$ despite the better statistical fit. Both models track the seasonal peaks and troughs reasonably well, as seen in the actual versus predicted time series plots (Fig \ref{fig:ts_comparison}), but don't fully capture the extreme spikes.

\begin{figure}[htbp]
    \centering
    \includegraphics[width=0.75\textwidth]{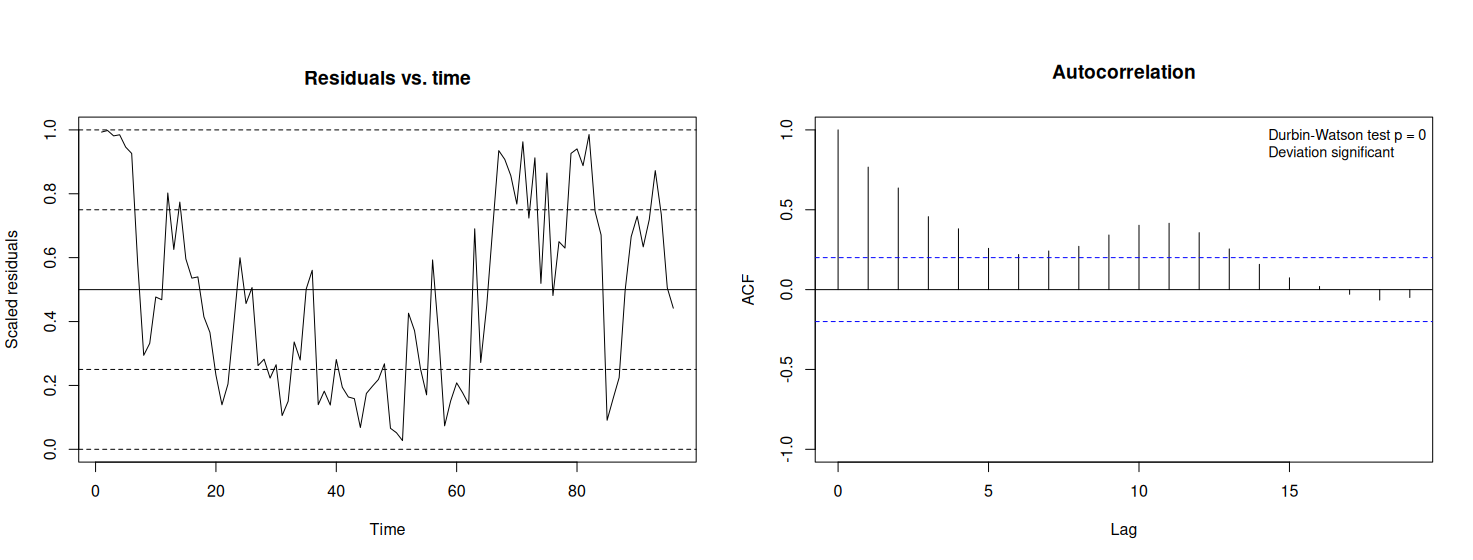}
    
    \vspace{1em}
    \includegraphics[width=0.75\textwidth]{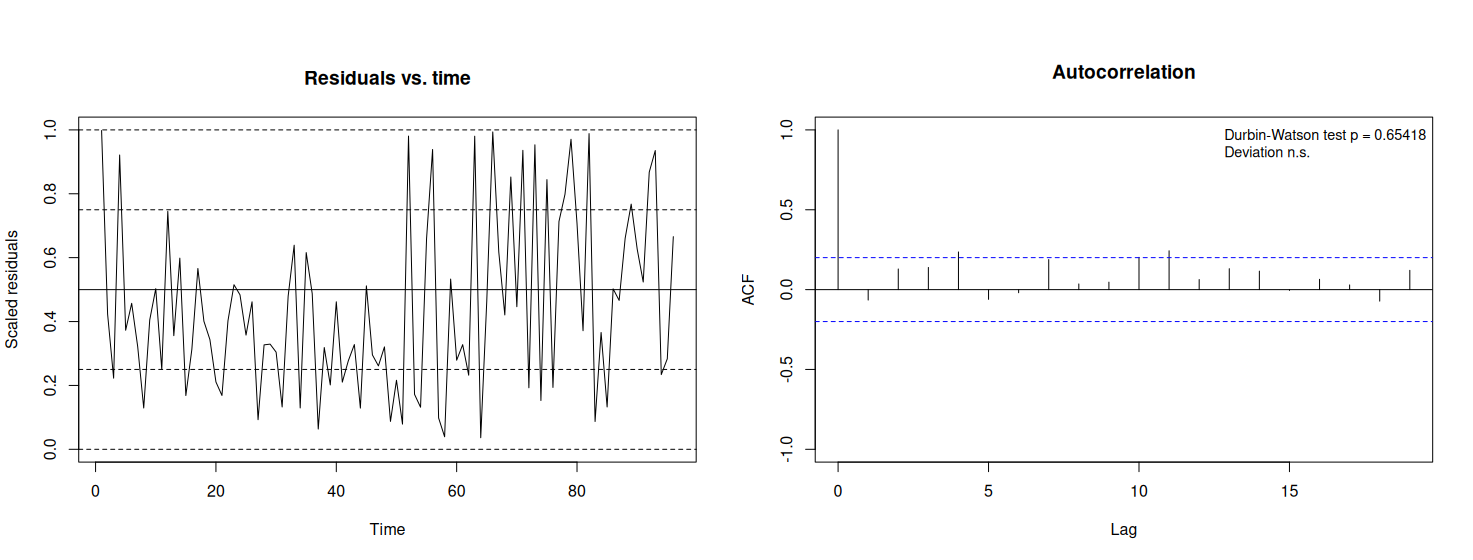}
    \caption{{\bf Temporal autocorrelation analysis of the randomized quantile residuals.} (Top) Autocorrelation diagnostics for the baseline negative binomial GLM. (Bottom) Autocorrelation diagnostics after applying the GLARMA Negative binomial framework.}
    \label{fig:acf_comparison}
\end{figure}

Although the baseline NB model correctly captured the heteroscedasticity, it left short-term month-to-month momentum unaddressed. As shown in Fig \ref{fig:acf_comparison} (Top), the Durbin-Watson test yielded a highly significant result ($p \approx 0$), and the Autocorrelation Function (ACF) plot had multiple spikes. The plot of scaled residuals versus time displayed a distinct, non-random wave pattern as well. This was resolved by fitting the GLARMA Negative binomial model, and the residual analysis was repeated. As seen in Fig \ref{fig:acf_comparison} (Bottom), the updated ACF plot is devoid of any spikes, confirming that the residual autocorrelation has been reduced to within the thresholds. The Durbin-Watson test is no longer significant ($p = 0.654$), and the scaled residuals over time now exhibit the random scatter characteristic of independent noise. Our progressive diagnosis confirms that the GLARMA Negative binomial model is the most appropriate framework here. 

\begin{figure}[!ht]
    \centering
    \includegraphics[width=0.75\textwidth]{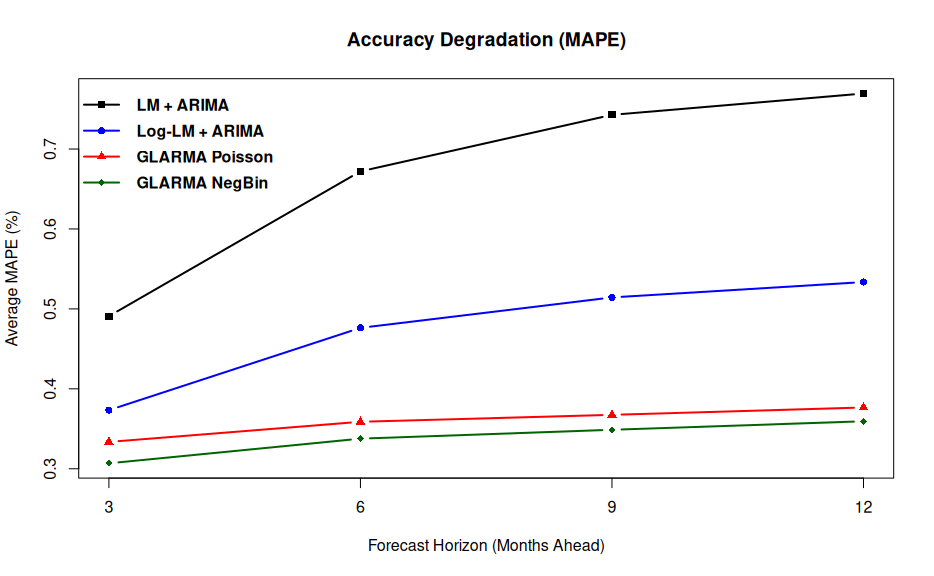}
    \caption{{\bf Accuracy degradation across forecast horizons.} Average Mean Absolute Percentage Error (MAPE) plotted against forecasting horizons of 3, 6, 9, and 12 months.}
    \label{fig:mape_degradation}
\end{figure}

With the model's structural adequacy verified in-sample, out-of-sample predictive performance across the four candidate frameworks was evaluated. The MAPE values and their standard deviations across all cross-validation folds are summarized in Table \ref{table:mape_results}. The models accumulate errors as the forecasting horizon expands, leading to a degradation in forecasting accuracy. Fig \ref{fig:mape_degradation} illustrates the average MAPE for each model across all four forecasting horizons. The MAPE values of LM + ARIMA degrade rapidly, with its average error inflating from 0.4909 at 3 months to 0.7695 at 12 months. The Log-LM + ARIMA is marginally better but still suffers significant accuracy loss. In contrast, both GLARMA models maintain a flat and stable trajectory even as the forecast pushes 12 months into the future. Ultimately, the GLARMA Negative binomial model consistently outperforms all other models, achieving the lowest error metrics at each horizon.

\begin{table}[!ht]
\begin{adjustwidth}{-2.25in}{0in}
\centering
\caption{\textbf{Out-of-sample predictive accuracy across forecast horizons.} Mean Absolute Percentage Error (MAPE $\pm$ Standard Deviation) evaluated via time series cross-validation.}
\begin{tabular}{|l+c|c|c|c|}
\hline
\multicolumn{1}{|l|}{\textbf{Horizon}} & \multicolumn{1}{|c|}{\textbf{LM + ARIMA}} & \multicolumn{1}{|c|}{\textbf{Log-LM + ARIMA}} & \multicolumn{1}{|c|}{\textbf{GLARMA Poisson}} & \multicolumn{1}{|c|}{\textbf{GLARMA NegBin}} \\ \thickhline
3 Months & 0.4909 $\pm$ 0.5607 & 0.3733 $\pm$ 0.3889 & 0.3336 $\pm$ 0.2335 & 0.3068 $\pm$ 0.1446 \\ \hline
6 Months & 0.6721 $\pm$ 0.7546 & 0.4763 $\pm$ 0.4158 & 0.3587 $\pm$ 0.1525 & 0.3376 $\pm$ 0.0994 \\ \hline
9 Months & 0.7428 $\pm$ 0.6549 & 0.5143 $\pm$ 0.3363 & 0.3674 $\pm$ 0.1202 & 0.3487 $\pm$ 0.0866 \\ \hline
12 Months & 0.7695 $\pm$ 0.5009 & 0.5335 $\pm$ 0.2688 & 0.3766 $\pm$ 0.0889 & 0.3594 $\pm$ 0.0649 \\ \hline
\end{tabular}
\label{table:mape_results}
\end{adjustwidth}
\end{table}

\begin{figure}[H]
    \centering
    \includegraphics[width=0.79\textwidth]{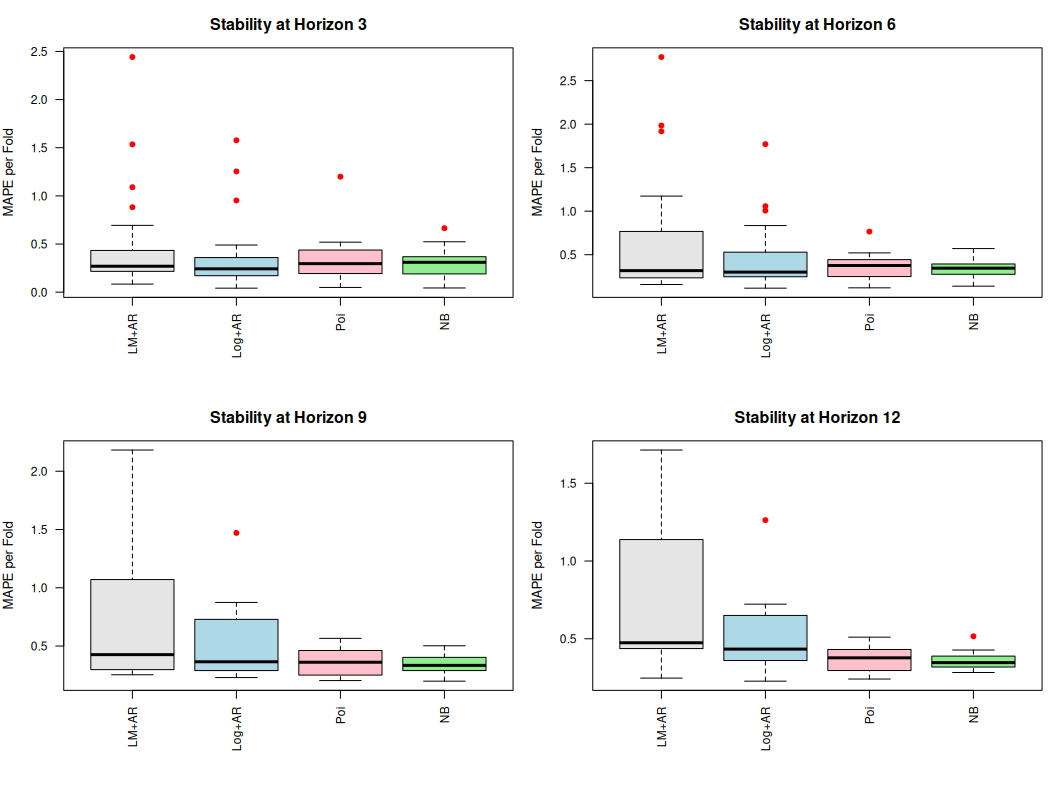}
    \caption{{\bf Forecast stability across cross-validation folds.} Boxplots displaying the distribution of Mean Absolute Percentage Error (MAPE) per fold at horizons 3, 6, 9, and 12.}
    \label{fig:stability_boxplots}
\end{figure}

Outside of average accuracy, the reliability of forecasts also depends on how stable they are. We analyzed the distribution of MAPE values across all cross-validation windows to evaluate this robustness. As visualized in the boxplots in Fig \ref{fig:stability_boxplots}, the LM + ARIMA and Log-LM + ARIMA models exhibit large variance and outliers, i.e., their predictions are volatile and dependent on the specific training window. In sharp contrast, the GLARMA Poisson and GLARMA Negative binomial models exhibit narrow interquartile ranges and few outliers. The GLARMA Negative binomial model has the lowest standard deviations (e.g., $\pm 0.0649$ at 12 months, compared to $\pm 0.5009$ for LM + ARIMA) and is visually the most stable. This shows that the GLARMA Negative binomial model is both the most structurally adequate in-sample and also the most accurate and reliable out-of-sample.

\section*{Conclusion}
This study demonstrates that monthly malaria incidence in Mumbai exhibits characterized dispersion and temporal dependence, rendering conventional Poisson-based approaches insufficient for reliable modeling and forecasting. While the negative binomial model successfully addressed the data's variance, residual diagnostics indicated significant autocorrelation, requiring further modeling. Incorporating an autoregressive component through the GLARMA framework produced statistically adequate models and improved forecasting performance across multiple validation periods. The results highlight the importance of jointly addressing distributional and temporal characteristics when analyzing epidemiological count data. The GLARMA Negative binomial framework provides a robust and interpretable approach for malaria modeling and forecasting, with potential applications in other infectious disease monitoring systems where accurate short-term prediction is essential for public health decision-making.

\section*{Funding}
Funding for this study was provided by the Gates Foundation (INV-044445).

\section*{Data availability}
The data and code that support the findings of this study are available on GitHub at \url{https://github.com/sommeraxe/count-ts}.

\section*{Disclaimer}
The work/opinion is based on research findings by the authors and not the opinion of the government.

\nolinenumbers

% Please compile your BiBTeX database using the "plos2025.bst" BibTeX style.
% This file is part of the current package.
% A sample BibTeX file is also included as "plos_bibtex_sample.bib".
%
% or
%
% Type in your references following Vancouver style and reference formatting instructions
% available at https://journals.plos.org/plosone/s/submission-guidelines#loc-references
% \begin{thebibliography}{}
% Text
% \end{thebibliography}

\bibliography{plos_bibtex_sample}

\end{document}